\documentstyle[aps]{revtex}
\begin{document}
\newcommand{\beq}{\begin{equation}}
\newcommand{\eeq}{\end{equation}}
\newcommand{\beqn}{\begin{eqnarray}}
\newcommand{\eeqn}{\end{eqnarray}}
\newcommand{\bmath}{\begin{mathletters}}
\newcommand{\emath}{\end{mathletters}}
%\draft
\twocolumn[\hsize\textwidth\columnwidth\hsize\csname @twocolumnfalse\endcsname
\title{Consequences of charge imbalance in superconductors within the theory 
\newline of hole superconductivity}
\author{J. E. Hirsch }
\address{Department of Physics, University of California, San Diego\\
La Jolla, CA 92093-0319}
 
\date{\today} 
\maketitle 
\begin{abstract} 
The theory of hole superconductivity proposes that the fundamental asymmetry
between electrons and holes in solids is responsible for superconductivity.
Here we point out a remarkable consequence of this theory:
 a tendency for negative charge to be expelled 
from the bulk of the superconductor towards the surface.
Experimentally observable consequences of this physics are discussed.
\end{abstract}
\vskip2pc]

\section{Introduction}

The theory of hole superconductivity\cite{hole1,hole2} asserts that superconductivity
originates in the fundamental charge asymmetry of condensed matter. This
asymmetry has its root in the fact that the positively charged proton is
far heavier than the negatively charged electron. The theory asserts that
only (positively charged) hole carriers in a solid can give rise to
superconductivity. Here we propose that a generic consequence of this
theory is that superconductors will have the tendency to expel negative charge
from their bulk. The negative charge will move to the surface, and the 
charge distribution in all superconductors will look qualitatively as in
Figure 1.

In this theory, superconductivity occurs through 'undressing' of heavy hole
carriers in electronic energy bands that are almost full\cite{undr}. The holes
are dressed due to the electron-electron interaction, which makes them heavy,
and the dressing is postulated to be an increasing function of electronic
band filling. When holes pair, they partially undress; the superfluid
carriers have smaller effective mass, and the associated lowering of kinetic
energy provides the superconducting condensation energy\cite{kine}.

The gap function has a slope of
universal sign\cite{hole2}, shown schematically in Figure 2 . This universal slope
reflects the electron-hole asymmetry that gives rise to superconductivity. The
quasiparticle excitation energy is given by\cite{hole2}
\beq
E_k=\sqrt{(\epsilon_k-\mu)^2+\Delta_k^2}=\sqrt{a^2(\epsilon_k-\mu-\nu)^2+\Delta_0^2}
\eeq
and its minimum occurs not at $\epsilon_k=\mu$, as in ordinary BCS theory,
but instead at 
\beq
\epsilon_k=\mu+\nu \equiv \mu '
\eeq
which defines a new chemical potential $\mu'$. The parameter $a$, the 
gap $\Delta_0$, and the shift in chemical potential $\nu=\mu '-\mu$ are
function of the microscopic parameters in the theory and can be calculated
as function of carrier concentration and temperature\cite{hole2}.

The coherence factors are given by their usual forms
\bmath
\beq
u_k^2=\frac{1}{2}(1+\frac{\epsilon_k-\mu}{E_k})
\eeq
\beq
v_k^2=\frac{1}{2}(1-\frac{\epsilon_k-\mu}{E_k})
\eeq
\emath
and because they are not symmetric around $\mu'$, quasiparticles carry
a net $positive$ charge $eQ^*$, given by
\beq
Q^*=\frac{2}{N}\sum_k(u_k^2-v_k^2) f(E_k)
\eeq
with $f$ the Fermi function. We have pointed out before that this charge
asymmetry will manifest itself in a universal asymmetry in NIS tunneling\cite{hole2}
(larger current for negatively biased superconducting electrode) and
in a $positive$ thermoelectric power of NIS tunnel junctions with a
temperature gradient across the juction\cite{thermo}.
As a consequence of quasiparticles being $positively$ charged, the
condensate will acquire an extra $negative$ charge.

Hence the superconductor is characterized by having two different
'chemical potentials'. The chemical potential $\mu$ corresponds to
the condensate, and $\mu'=\mu+\nu$ to the quasiparticle excitations. 
In a $hole$ representation, $\mu' >\mu$ (in an electron representation, $\mu >\mu'$).
The negatively charged condensate, by virtue of being a superfluid as well
as because of the effective mass reduction that occurs due to undressing,
is highly mobile, in contrast to the quasiparticles which experience normal
scattering and have the higher effective mass characteristic of the
normal state dressed carriers. As a consequence, we expect that the condensate will have a 
tendency to move $out$ of the bulk of the superconductor, so as to
tend to equate the chemical potentials $\mu$ and $\mu'$ in the bulk. Because of
overall charge neutrality, the negative charge will accumulate near the
surface of the superconductor, giving rise to the qualitative 
charge distribution 
shown in Figure 1. 

An estimate of the maximum amount of charge that will be expelled from the
bulk of the superconductor is given by the ratio of the difference in
chemical potentials to the bandwidth $D$
\beq
n_{max}=\frac{2(\mu'-\mu)}{D}=\frac{2\nu}{D}
\eeq
carriers per site. For typical parameters in the model $\nu\sim1meV$, $D\sim0.5 eV$, 
yielding $n_{max}\sim 0.004$ electrons per atom. However the tendency to 
charge expulsion will be counteracted by Coulomb charging energy. The net
effect will be most pronounced in superconducting bodies of small
volume, i.e. large surface to volume ratio.

We next discuss some expected experimental consequences of this effect:

(1) Consider a superconductor in the mixed state. Depending on the value of
the applied magnetic field, the surface to volume ratio of the superconducting
regions can be made very large. We expect that negative charge will be expelled from
the superconducting regions into the normal vortex cores. Recently it has
been reported\cite{nqr} that indeed in $YBa_2Cu_3O_7$ in the mixed state,
the vortices are negatively charged relative to the superconducting
regions. However, reference 6 also reports that in $YBa_2Cu_4O_8$ the 
vortices are $positively$ charged, and concludes that generally vortices
will be negatively charged in the overdoped regime and positively
charged in the underdoped regime. We believe that this conclusion is
erroneous: our theory predicts universally negatively charged vortices.
The conclusion of Ref. 6 for $YBa_2Cu_4O_8$ could be due either to an
experimental artifact or to other non-intrinsic physical processes
occuring in that material.

(2) Consider the recent remarkable experiment reported by
 Tao and coworkers\cite{balls}: granular superconducting particles in
an electric field aggregate to form round balls of macroscopic
dimensions. We propose the following qualitative explanation : in the 
absence of electric field, the granular particles with charge
distribution as in Figure 1 produce no electric field outside, hence there is
no electrostatic force between granules. When the electric field is turned
on the granules polarize and the resulting electric dipoles attract,
so that granules will join and allign parallel to the field direction.
However when  granules join the negative charge on the parts of the 
surfaces that join will move to the outer surface. The resulting
negative charge distribution on the surface will exert a force to
distort the elongated shape onto a spherical shape which gives rise to
minimum surface energy of the negative charge. The result is the
spherical ball of macroscopic dimensions seen experimentally\cite{balls}.

(3) In polycrystalline samples of high $T_c$ materials it is observed that
 a depletion of hole carriers occurs close to grain
boundaries\cite{grain1}. In fact it was recently proposed that overdoping
of grain boundaries is an effective way to increase critical 
current densities in high $T_c$ materials\cite{grain2}. We propose that this
depletion of hole carriers occurs because  negative charge expelled
from the bulk of the superconductor will migrate to the grain
boundaries and plug holes, thus underdoping those regions. The direct
observation that grain boundaries are underdoped in $YBCO$\cite{grain1} was made at
liquid nitrogen temperatures\cite{grain3}, hence in the superconducting
state. Our scenario predicts that the effect should disappear at
temperatures above $T_c$.

(4) If there is negative surface charge in superconductors one would 
expect that it will be relatively easy to remove when it is rubbed against
a non-superconductor which is poorly or non-conducting (otherwise the
charge would be transfered back upon separation). Hence superconductors
should be at the top of the triboelectric series\cite{tribo}. To our knowledge,
triboelectric properties of superconductors have never been examined.

To obtain quantitative answers for the charge distribution in a superconductor
of given shape, or in the mixed state, it is necessary to solve the
Bogoliubov - de Gennes equations, which will be the subject of future
work. We expect that quite generally a higher density of negatively charged
superfluid near the surfaces will result. Note also that in a superconducting
body the electric current is carried only by the superconducting electrons
in a surface layer, of thickness given by the penetration depth. It is
natural to expect that the higher density of superfluid should occur in the
region where it is useful to conduct electricity.

In conclusion, we point out that the inhomogeneous charge distribution of
figure 1 mirrors the one at the atomic level: in an atom, the heavy
positive charge is concentrated in the nucleus and the light negative 
charge extends out. Within the theory of hole superconductivity,
superconductors are solids where many $antibonding$ electronic states
are occupied. To relieve the associated energy cost these
electrons, which have high kinetic energy and effective mass, condense into 
the superconducting state, thus lowering their kinetic energy and 
effective mass. The resulting light superfluid moves out towards the surface,
just as light electrons in atoms do not remain confined within the 
dimensions of the positive nucleus. Thus we propose that superconducting
solids can in some sense be regarded as 'giant atoms'. Futher
consequences of this physics will be discussed in future work.

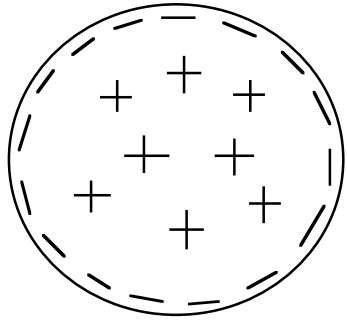
\begin{figure}
\caption { 
Schematic picture of a spherical superconducting body. Negative
charge is expelled from the bulk to the surface.}
\label{Fig. 1}
\end{figure}
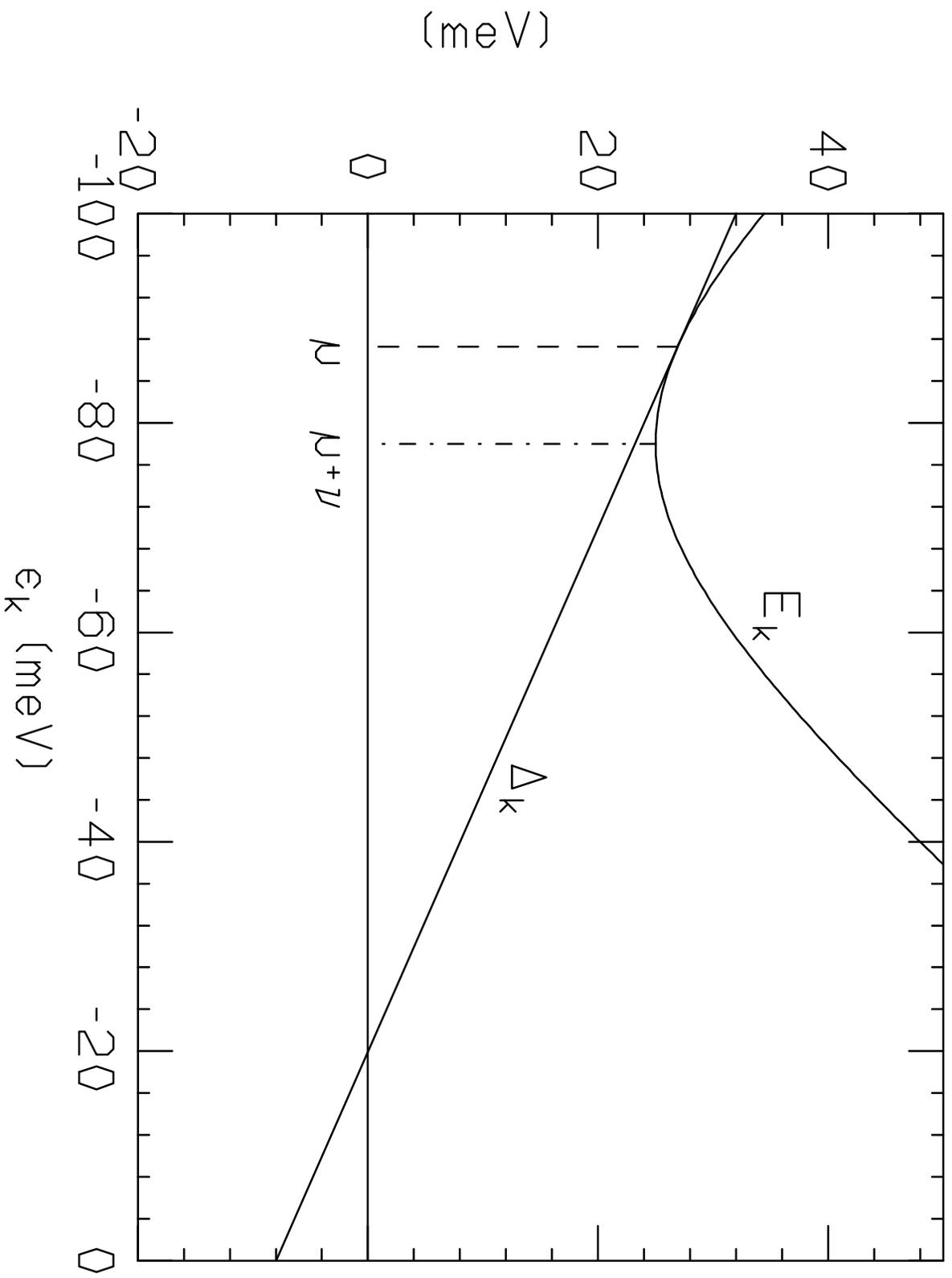
\begin{figure}
\caption { 
Schematic picture of the energy gap function $\Delta _k$ and
quasiparticle energy $E_k$ versus hole kinetic energy $\epsilon_k$
in the model of hole superconductivity. The minimum in the
quasiparticle energy is shifted from the chemical potential $\mu$ to
$\mu + \nu$.}
\label{Fig. 2}
\end{figure}
\end{document}